\documentclass[fleqn,usenatbib]{mnras}
\usepackage{newtxtext,newtxmath}
\usepackage[normalem]{ulem}

\usepackage[T1]{fontenc}
\usepackage{ae,aecompl}

\usepackage{graphicx}	
\usepackage{amsmath}	
\usepackage{amssymb}	
\usepackage{url}

\title[Primordial magnetic fields and EDGES measurements]{Primordial magnetic fields during the cosmic dawn in light of EDGES 21-cm signal}

\author[A. Bera et al.]{
Ankita Bera,$^{1}$\thanks{E-mail:ankita1.rs@presiuniv.ac.in}
Kanan K. Datta,$^{1}$\thanks{E-mail: kanan.physics@presiuniv.ac.in}
Saumyadip Samui$^{1}$\thanks{E-mail: saumyadip.physics@presiuniv.ac.in}
\\
$^{1}$Department of Physics, Presidency University, 86/1 College Street, Kolkata, 700073, India}

\date{Accepted XXX. Received YYY; in original form ZZZ}

\pubyear{2020}

\begin{document}
\label{firstpage}
\pagerange{\pageref{firstpage}--\pageref{lastpage}}
\maketitle

\begin{abstract}
We study prospects of constraining the primordial magnetic field (PMF) and its evolution during the dark ages and cosmic dawn in light of EDGES 21-cm signal. Our analysis has been carried out on a `colder IGM' background which is one of the promising avenues to interpret the EDGES signal. We consider the dark matter-baryon interactions for the excess cooling. We find that the colder IGM suppresses both the residual free electron fraction and the coupling coefficient between the ionised and neutral components. The Compton heating also gets affected in colder IGM background. Consequently,  the IGM heating rate due to the PMF enhances compared to the standard scenario.
Thus, a significant fraction of the magnetic energy, for $B_0 \lesssim 0.5 \, {\rm nG}$, gets transferred to the IGM and the magnetic field decays at much faster rate compared to the simple $(1+z)^2$ scaling during the dark ages and cosmic dawn. This low PMF is an unlikely candidate for explaining the rise of the EDGES absorption signal at lower redshift.  We also see that the PMF and DM-baryon interaction together introduces a plateau like feature in the redshift evolution of the IGM temperature. 
We find that the upper limit on the PMF depends on the underlying DM-baryon interaction. Higher PMF can be allowed when the interaction cross-section is higher and/or the DM particle mass is lower. Our study shows that the PMF with $B_0$ up to  $\sim 0.4 \, {\rm nG}$, which is ruled out in the standard model, can be allowed if  DM-baryon interaction with suitable cross-section and DM mass is considered. 

\end{abstract}

\begin{keywords}
cosmology: dark ages, reionization, first stars -- cosmology: dark matter -- magnetic fields -- methods: analytical
\end{keywords}



\section{Introduction}
The global redshifted HI 21-cm signal from the dark ages and cosmic dawn is a promising tool to study the primordial magnetic field \citep[see][for a review]{kandu2016}. The primordial magnetic field can heat up the Hydrogen and Helium gas in the inter galactic medium (IGM) by processes such as the ambipolar diffusion (AD) and decaying turbulence (DT) \citep{jedamzik98, kandu98, KK14, Chluba15}. This indirectly affects the spin temperature and the globally averaged redshifted HI 21-cm  signal \citep{sethi05}. Furthermore,  growth of structures during the cosmic dawn gets accelerated in  presence of magnetic field in the IGM. As a consequence, the primordial magnetic field can have an important impact on the formations of the first luminous sources \citep{sethi08, Schleicher08}. A substantial amount of theoretical work has been carried out to understand, in detail, the role of the primordial field on the HI 21-cm signal \citep{tashiro2006, Schleicher09, Venumadhav2017, kunze2019}, early structure formation during the cosmic dawn and reionization \citep{kim1996, yamazaki2006, pandey2015}.

The measurements of the global HI 21-cm absorption signal by the EDGES experiments in the redshift range $z \sim 14$ to $20$ \citep{EDGES18} have opened up a possibility to constrain the primordial magnetic field and understand its evolution during the cosmic dawn and dark ages. In a recent work, \citet{Minoda19} has exploited the EDGES data to put an upper limit on the primordial magnetic field. The analysis has been carried out on the backdrop of the standard cosmological model and baryonic interactions of the IGM. However, the measured EDGES absorption signal is  $\sim 2 -3$ times stronger compared to predictions by the standard model. If the measurements are confirmed, one promising way to explain the measured signal is to consider the IGM to be significantly  `colder' compared to the IGM kinetic temperature predicted by the standard scenario. Thus, one needs to consider a non-standard cooling mechanism such as the DM-baryon interaction in order to make the IGM colder \citep{Tashiro14, Munoz15}. This avenue has been widely explored to explain the unusually strong absorption signal found by the EDGES experiment \citep[see e.g.,][]{Barkana18Nature, Barkana18PRD, Munoz18, Munoz18a}. 

Constraints on the primordial magnetic field using the global 21-cm absorption signal in the colder IGM background would, in principle, be different from constraints obtained in the standard scenario. Because, colder IGM enhances the Hydrogen recombination rate which, in turn, reduces the residual free electron fraction during the dark ages and cosmic dawn \citep{Datta20}. In addition, the coupling between the ionized and neutral component which has direct impact on the IGM heating also get suppressed in the 'colder IGM' scenario. Moreover, the heating rate due to the Compton process, which depends on the IGM kinetic temperature and the residual free electron fraction, too gets affected when the background IGM temperature is lower. Together all these effects enhance the IGM heating rate due to the primordial magnetic field. Consequently, small amount of magnetic field would be enough to keep IGM temperature at a certain label. On the contrary, significantly more magnetic energy would be transferred to the IGM due to the enhanced heating rate which would affect the redshift evolution of the primordial magnetic field itself and the heating at later reshifts. If one considers DM-baryon interaction in order to make the IGM colder, the exact constraints on primordial magnetic field should also depend on the mass of the DM particles and the interaction cross section between the DM particles and baryons. Thus, it is important to highlight these aspects in order to understand the role of the primordial magnetic field on the 21-cm absorption signal and put limits on primordial magnetic field. Recently, \citet{bhatt2020} has used the EDGES low band measurements to study constraints on the primordial magnetic field in presence DM-baryonic interaction. Various other observations such the CMBR, the Sunyaev-Zel'dovich effect, the star formation, blazar light curve have been exploited to constrain the primordial magnetic field \citep{planck15b, saga20, Minoda17, marinacci2016, takahashi2013}. Constraining the primordial magnetic field is very important as it can shed light on its origin and evolution. 

In this work, we study the constraints on the primordial magnetic field using the EDGES 21-cm absorption profile on the backdrop of the colder IGM scenario. We consider interactions between cold DM particles and baryons \citep{Tashiro14, Munoz15} which makes the IGM colder as compared to that in the standard predictions. In addition, we study the redshift evolution of the primordial magnetic field during dark ages and cosmic dawn. The differential brightness temperature in the EDGES absorption profile starts increasing at redshift $z \sim 16$ which suggests that heating of the IGM started around that redshift.  Here, we also investigate if the primordial magnetic heating is able to explain this behavior. Our analysis also allows us to study the constraints on the mass of the DM particle and the interaction cross section in presence of the primordial magnetic field. 
  
The structure of the paper is as follows. A brief discussion and essential equations regarding the redshifted HI 21-cm signal, dark matter - baryon interaction, heating due to the primordial magnetic field and redshift evolution of IGM temperature in presence of both the primordial magnetic field and DM-baryon interaction are presented in subsections 2.1, 2.2, 2.3 and 2.4 respectively. We discuss our results in section~3 and present summary and discussion in section~4. Throughout our work we use cosmological parameters $\Omega_{m0}= 0.3$, $\Omega_{b0} = 0.0486$, $h=0.677$, $\Omega_{\Lambda 0} = 0.7$ consistent with the Plank measurements \citep{planck15}. 


\section{HI 21-cm signal in colder IGM}
\label{sec:formulation}
\subsection{HI 21-cm signal}
\label{subsec:HI signal}
The globally averaged differential brightness temperature corresponding to the redshifted HI 21-cm at redshift $z$ can be written as \citep{bharadwaj05, furlanetto06},
\begin{eqnarray}
    \frac{T_{21}}{\rm mK} = 27 x_{\rm HI} \left(1-\frac{T_\gamma}{T_s}\right) \left(\frac{\Omega_{\rm b0} h^2}{0.02} \right) \left(\frac{0.15}{\Omega_{\rm m0} h^2} \right)^{0.5} \left( \frac{1+z}{10} \right)^{0.5},
\end{eqnarray}
where $T_\gamma$ and $x_{HI}$ are the CMBR temperature and neutral Hydrogen fraction respectively.  The spin temperature $T_{\rm s}$ which is a measure of  population ratio of the ground state Hydrogen atoms in the triplet and singlet states is defined as,
\begin{equation}
    \frac{n_1}{n_0} = \frac{g_1}{g_0} \exp{(-T_\ast/T_{\rm s})},
\end{equation}
where $n_0$ and $n_1$ are the number densities of ground state Hydrogen atoms in the singlet and triplet states respectively, and $g_0=1$ and $g_1=3$ are the degeneracies of these states. Further, $T_\ast = h_p \nu_e/k_B = 0.068  $~K is the characteristic temperature corresponding to the HI 21-cm transition.  The Ly-$\alpha$ photons emitted from the very first stars/galaxies help the spin temperature $T_{\rm s}$ to couple with the IGM kinetic temperature $T_g$. Since we are interested in the Ly-$\alpha$ saturated part of the EDGES absorption profile, we assume $T_s=T_g$ for the rest of the paper.

\subsection{Dark matter-baryon interaction}
\label{subsec:dark-matter}
Interactions between the cold DM particles and baryons are expected to help the IGM to cool faster than the standard adiabatic cooling and can explain the unusually strong absorption signal found by the EDGES experiments \citep{Barkana18Nature}. We consider Rutherford like velocity dependent interaction cross section which is modelled as  $\sigma = \sigma_0 (v/c)^{-4}$.  The milli-charged dark matter model follows this kind of interaction and is a potential candidate for explaining the EDGES trough \citep{Munoz18, Munoz18a}. Here we adopt the DM-baryon interaction model presented in \citet{Munoz15}. The cooling rate of baryon due to such interaction is modelled as, 
\begin{eqnarray}
    \frac{dQ_b}{dt} = \frac{2 m_b \rho_{\chi} \sigma_0 e^{-r^{2} / 2} (T_{\chi} - T_g) k_B c^4}{(m_b + m_{\chi})^2 \sqrt{2\pi} u^{3}_{th}} \nonumber \\
    + \frac{\rho_{\chi}}{\rho_m} \frac{m_{\chi} m_b}{m_{\chi} + m_b} V_{\chi b} \frac{D(V_{\chi b})}{c^2}.
    \label{eq:dQ_b}
\end{eqnarray}
Similarly, the heating rate of the DM, $\Dot{Q_\chi}$ can be obtained by just replacing $b \leftrightarrow \chi$ in the above expression due to symmetry.  Here, $m_\chi$, $m_b$ and $\rho_\chi$, $\rho_b$ are the masses and energy densities of dark matter and baryon respectively.  We can see from equation~(\ref{eq:dQ_b}) that the heating rate is proportional to the temperature difference between two fluids i.e. $(T_{\chi} - T_g)$. The second term in equation~(\ref{eq:dQ_b}) arises due to the friction between dark matter and baryon fluids as they flow at different velocities. Hence both the fluids get heated up depending on their relative velocity $V_{\chi b}$ and the drag term $D(V_{\chi b})$ given as,
\begin{equation}
\frac{dV_{\chi b}}{dz} = \frac{V_{\chi b}}{1+z} + \frac{D(V_{\chi b})}{H(z)(1+z)}
\label{eq:vxb}
\end{equation}
and
\begin{equation}
D(V_{\chi b})  = \frac{\rho_m \sigma_0 c^4}{m_b + m_\chi} \frac{1}{V^2_{\chi b}} F(r).
\end{equation}
The variance of the thermal relative motion of dark matter and baryon fluids $u_{th}^2 = k_B(T_b/m_b + T_\chi/m_\chi$)  and $r =V_{\chi b}/u_{th}$. The function $F(r)$ is given by
\begin{equation}
F(r) = erf \Big( \frac{r}{\sqrt{2}} \Big) - \sqrt{ \frac{2}{\pi}} r e^{-r^2/2}.
\end{equation}
We see that $F(r)$ grows with r, $F(0) = 0$ when $r = 0$ and $F(r) \rightarrow 1$ when $r \rightarrow \infty$. This ensures that the heating due to the friction is negligible when the relative velocity $V_{\chi b}$ is smaller compared to the thermal motion of dark matter and baryon fluid $u_{th}$. However, it can be significant if $V_{\chi b}$ is higher than $u_{th}$.

\subsection{IGM heating due to primordial magnetic field}
\label{subsec:PMF} 
Magnetic field exerts Lorentz force on the ionized component of the IGM. This causes rise in the IGM temperature, $T_g$. There are mainly two processes namely the ambipolar diffusion (AD) and decaying turbulence (DT) by which the magnetic field can heat up the IGM during the cosmic dawn and dark ages. We follow the prescription presented in \citet{SS05} and \citet{Chluba15} to calculate the rate of heating due to these two processes.  The heating rate (in unit of energy per unit time per unit volume)  due to the ambipolar diffusion is given by,
\begin{eqnarray}
    \Gamma_{\rm AD} = \frac{(1-x_e)}{\gamma x_e \rho_b^2} \frac{\Big \langle \vert (\nabla\times \boldsymbol{B})\times \boldsymbol{B} \vert^2 \Big \rangle}{16\pi^2},
    \label{eq:gamma_AD}
\end{eqnarray}
where $x_e = n_e/n_{\rm H}$ is the residual free electron fraction and $n_{\rm H} = n_{\rm HI} + n_{\rm HII}$. We assume $n_{\rm HII} = n_e$ as Helium is considered to be fully neutral in the redshift range of our interest. Further, $\rho_b$ is the baryon mass density at redshift $z$, and the coupling coefficient between the ionized and neutral components is $\gamma=\langle \sigma v \rangle _{HH^+}/{2m_H}= 1.94 \times 10^{14} \, (T_g/{\rm K})^{0.375} \, {\rm cm}^3 {\rm gm}^{-1} {\rm s}^{-1}$. The Lorentz force can be approximated as $ \Big \langle \vert (\nabla\times \boldsymbol{B})\times \boldsymbol{B} \vert^2 \Big \rangle \approx 16 \pi^2 \, \rho_B(z)^2 \, l_d(z)^{-2} f_L(n_B+3)$ \citep{Chluba15}, where $\rho_B(z)=|{\bf B}|^2/8\pi$ is the magnetic field energy density at redshift $z$, $f_L(p)=0.8313[1-1.02 \times 10^{-2} p]p^{1.105}$, and  $l_d^{-1}= (1+z) \,k_D$. The damping scale is given by $k_D \approx 286.91 \, (B_0/{\rm nG})^{-1}\, {\rm Mpc}^{-1}$ \citep{KK14}. We note that the above heating rate is inversely proportional to the coupling coefficient  $\gamma$ and the residual electron fraction $x_e$. Furthermore, both $\gamma$ and the ionization fraction, $x_e$ gets suppressed when the IGM is colder compared to the standard scenario. As a result, the ambipolar heating rate becomes more efficient during the cosmic dawn and dark ages.

The heating rate due to the  decaying turbulence is described by,
\begin{eqnarray}
    \Gamma_{\rm DT} = \frac{3m}{2} \frac{\Big\lbrack \ln{\left(1+\frac{t_i}{t_d} \right)}\Big\rbrack ^m}{\Big \lbrack \ln{\left(1+\frac{t_i}{t_d}\right)} + \frac{3}{2} \ln{\left(\frac{1+z_i}{1+z}\right)}\Big \rbrack^{m+1}} H(z)\, \rho_B(z),
    \label{eq:gamma_DT}
\end{eqnarray}
where $m=2 \, (n_B+3)/(n_B+5)$, and $n_B$ is the spectral index corresponding to the primordial magnetic field.  The physical decay time scale ($t_d$) for turbulence and the time ($t_i$) at which decaying magnetic turbulence becomes dominant are related as $t_i/t_d \simeq 14.8(B_0/{\rm nG})^{-1}(k_D/{\rm Mpc}^{-1})^{-1}$ \citep{Chluba15}. The heating rate due to the decaying turbulence is more efficient at early times as it is proportional to the Hubble rate, $H(z)$ and the primordial magnetic energy density. The effect monotonically decreases at lower redshifts and becomes sub-dominant during the cosmic dawn and dark ages. 

It is often assumed that, like the CMBR energy density,  the primordial magnetic field and energy density scale with redshift $z$ as $B(z) = B_0 (1+z)^2$ and $\rho_B(z) \sim (1+z)^4$ respectively under magnetic flux freezing condition.  However, the magnetic field energy continuously gets transferred to the IGM through the ambipolar diffusion and decaying turbulence processes. For the magnetic field with $B_0 \gtrsim 1 \, {\rm nG}$, the transfer may be insignificant compared to the total magnetic filed energy and the above scalings holds. However, this may not be a valid assumption for lower magnetic field $B_0 \lesssim  0.1 $ nG.  Therefore, we self-consistently calculate the redshift evolution of the magnetic field energy using the following equation,
\begin{equation}
    \frac{d}{dz} \left( \frac{\vert \boldsymbol{B} \vert^2}{8 \pi} \right) = \frac{4}{1+z} \left( \frac{\vert \boldsymbol{B} \vert^2}{8 \pi} \right) + \frac{1}{H(z)\,(1+z)} (\Gamma_{\rm DT} + \Gamma_{\rm AD}). 
    \label{eq:mag_density}
\end{equation}
The first term in the rhs quantifies the effect due to the adiabatic expansion of universe, and the second term quantifies the loss of the magnetic energy due to the IGM heating described above.

\subsection{Temperature evolution}
\label{subsec:temp and ion frac}
This section focuses on the evolution of the IGM kinetic temperature $T_g $ from the recombination epoch to the cosmic dawn. Considering the effects described in Section~\ref{subsec:dark-matter} and \ref{subsec:PMF}, the evolution of IGM gas temperature ($T_g$) can be written as,
\begin{eqnarray}
    \frac{dT_g}{dz} = \frac{2T_g}{1+z} - \frac{32\sigma_T\sigma_{SB} T_0^4}{3m_ec^2H_0\sqrt{\Omega_{m0}}}\left(T_\gamma-T_g\right)\left(1+z\right)^{3/2} \frac{x_e}{1+x_e} \nonumber \\
    - \frac{2}{3k_B H(z)\,(1+z) } \left [ \Dot{Q_b}+ \frac{\Gamma}{n_{\rm tot} } \right ].
	\label{eq:Tg}
\end{eqnarray}
The first two terms on the rhs describe the adiabatic cooling due to expansion of the universe and Compton heating due to interaction between CMBR and free electrons respectively. Further, $\Dot{Q_b}$ is the heating/cooling rate per baryon due to interactions between the DM particles and baryons (see eq.~\ref{eq:dQ_b}) and $\Gamma=\Gamma_{\rm AD}+\Gamma_{\rm DT}$ is the total rate of heating per unit volume due to the primordial magnetic field described in eqs.~\ref{eq:gamma_AD} and  \ref{eq:gamma_DT}. Also,  $ k_B$, $ \sigma_T$, $\sigma_{SB}$, $m_e$ are the Boltzman constant, Thomson scattering cross-section, Stefan Boltzman constant and the rest mass of an electron respectively. Further, $n_{\rm tot} \approx n_{\rm H}(1+f_{He}+x_e)$ denotes the total number density of baryon particles, and $n_{\rm H}$ is the number density of Hydrogen. Taking the Helium mass fraction $Y_P=0.24$, the fraction of Helium atoms with respect to hydrogen atoms, $f_{\rm He}$ becomes $0.079$. The evolution of the DM temperature $T_{\chi}$ can be calculated using,
\begin{equation}
    \frac{dT_\chi}{dz} = \frac{2T_\chi}{1+z} - \frac{2}{3k_B} \frac{\Dot{Q_\chi}}{H(z)\, (1+z)}.
    \label{eq:Tchi}
\end{equation}
The first and second terms on the rhs quantify the adiabatic cooling and heating rate per dark matter particle due to its interactions with baryons respectively. 
\begin{figure}
	\includegraphics[width=\columnwidth]{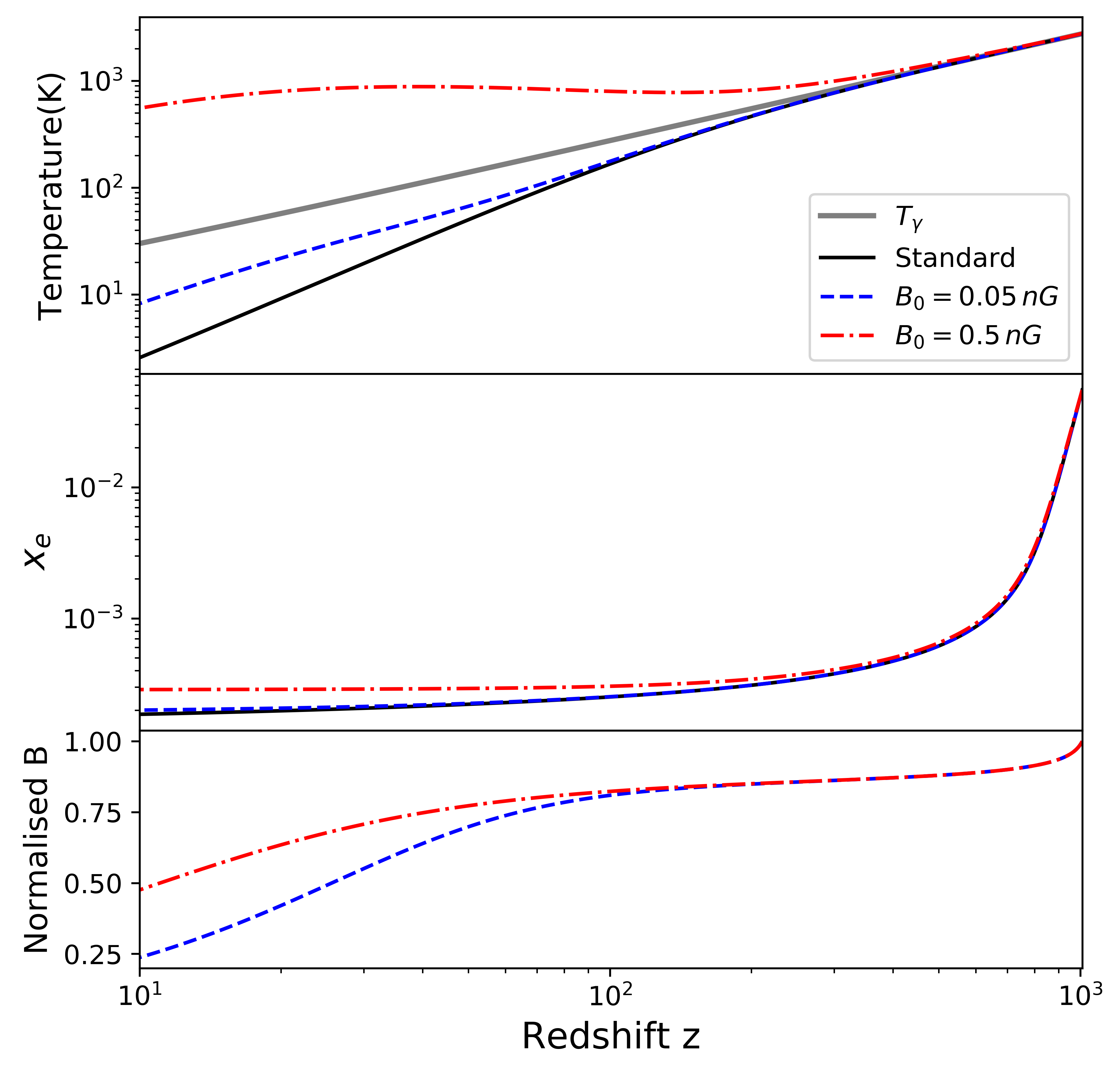}
    \caption{The upper and middle panels show the IGM kinetic temperature $T_g$ and residual free electron fraction $x_e$ as a function of redshift  in presence of the primordial magnetic field.  The lower solid (black), dashed-dotted (red) and dashed (blue) lines correspond to the primordial magnetic field with $B_0 = 0$,  $0.05$ and $0.5\, {\rm nG}$ respectively. The upper solid (black) line shows the CMBR temperature $T_{\gamma}$. We do not include the DM-baryon interaction here. The lower panel shows the normalised magnetic field i.e. $\frac{B(z)}{B_0 (1+z)^2}$ for the same $B_0$ values mentioned above.}
    \label{fig:PMF}
\end{figure}

We note that the residual free electron fraction $x_e$ influences the IGM heating through the Compton heating (eq.~\ref{eq:Tg}) and ambipolar diffusion (eq.~\ref{eq:gamma_AD}). We calculate the residual free electron fraction using the equation \citep{Peebles1968},
\begin{eqnarray}
    \frac{dx_e}{dz} = \frac{C}{H(z)\,(1+z)} \left[\alpha_e\,x_e^2 n_{\rm H} -\beta_e\,(1-x_e) \, e^\frac{-h_p\nu_\alpha}{k_BT_g}\right] \nonumber\\
    - \frac{\gamma_e\,n_H(1-x_e)x_e}{H(z)\,(1+z)},
    \label{eq:ion_frac}
\end{eqnarray}
where $\alpha_e(T_g)$, $\beta_e(T_{\gamma)}$ and $\gamma_e(T_g)$ are the recombination, photoionization and collisional ionization coefficients respectively. We note that $\alpha_e$ and $\gamma_e$ depend on the IGM temperature $T_g$. In contrast, $\beta_e$ depends on the CMBR temperature \citep[see][for a detailed discussion]{Chluba15}. For the recombination co-efficient we use $\alpha_e (T_g) = F \times 10^{-19} (\frac{a t^b}{1 + c t^d}) \hspace{0.1cm} {\rm m^3 s^{-1} } $, where $a = 4.309$, $b = -0.6166$, $c = 0.6703$, $d = 0.53$, $F = 1.14$ (the fudge factor)  and $t = \frac{T_g}{10^4 \, {\rm K}} $. Further, $\beta_e$ is calculated using the relations  $\beta_e (T_{\gamma}) = \alpha_e (T_{\gamma}) \Big( \frac{2 \pi m_e k_B T_\gamma}{h^2_p} \Big)^{3/2} e^{-E_{2s}/k_B T_\gamma}$ \citep{Seager1999,Seager2000}. The Peebles factor is given by $C = \frac{1+ K \Lambda (1-x) n_{H}}{1+K(\Lambda+\beta_{e})(1-x) n_{H}}$, where $\Lambda=8.3 \, {\rm s}^{-1}$ is the rate of transition from (hydrogen ground state) $2s\rightarrow1s$ state through decaying two photons. Further, $K=\frac{\lambda_{\alpha}^{3}}{8\pi H(z)}$, $\gamma_e(T_g) =0.291 \times 10^{-7} \times U ^{0.39} \frac{\exp(-U)}{0.232+U}  \, {\rm cm^3/s} $ \citep{Minoda17} with $h_{p}  \nu_{\alpha} = 10.2 \, {\rm eV}$ and $U=\vert E_{1s}/k_B T_g \vert$. 

We see from equations~(\ref{eq:dQ_b}) and (\ref{eq:Tg}), that the IGM temperature becomes velocity ($V_{\chi b}$) dependent as soon as  the dark matter-baryon interaction is taken into consideration which, in turn, modifies the brightness temperature $T_{21}$. Therefore, the observable global HI 21-cm brightness temperature is calculated by averaging over the velocity $V_{\chi b}$ as,
\begin{equation}
\langle T_{21} (z) \rangle = \int d^3 V_{\chi b} T_{21} (V_{\chi b}) P(V_{\chi b}),
\label{eq:average}
\end{equation}
where the initial velocity $V_{\chi b, 0}$ follows the probability distribution 
\begin{equation}
P(V_{\chi b, 0}) = \frac{e^{-3 V^2_{\chi b, 0}/( 2 V^2_{\rm rms})}}{(\frac{2 \pi}{3} V^2_{\rm rms})^{3/2}}.
\label{eq:probability}
\end{equation}
In order to calculate the velocity averaged IGM temperature $\langle T_g(z) \rangle$ and ionization fraction $\langle x_e \rangle$, the same procedure is followed.


\section{Results and Discussion}
\label{sec:results}
We simultaneously solve equations~(\ref{eq:dQ_b}), (\ref{eq:vxb}), (\ref{eq:mag_density}), (\ref{eq:Tg}), (\ref{eq:Tchi}) and (\ref{eq:ion_frac})  to evaluate $T_g$ and $x_e$ for a range possible values of the dark matter particle mass $m_{\chi}$, the interaction cross-section $\sigma_{45}=\frac{\sigma_0}{10^{-45} \, {\rm m^2}}$, and the initial magnetic field ($B_0$) for a given $V_{\chi b}$. We then use eq. \ref{eq:average} to calculate the averaged quantities such as $\langle T_{21} (z) \rangle$, $\langle T_g(z) \rangle$ and $\langle x_e \rangle$. Note that all values/results quoted below are these average quantities even if we don't mention them explicitly. Below we discuss our results on the heating due the primordial magnetic field, impacts of the DM-baryonic interaction in presence/absence of the magnetic field. In addition we study, in the context of the colder IGM background, the role of the residual free electron fraction $x_e$, evolution of the primordial magnetic field and the upper limit on the primordial magnetic field using EDGES absorption profile.  We set the following initial conditions at redshift $z_i=1010$: $T_{\rm gi}  = 2.725(1+z_i) \, {\rm K};\, T_{\chi i}= 0,\, V_{\chi b,i} = V_{\rm rms i} = 29 \, {\rm km/s}$, $B_i=B_0(1+z_i)^2$ and $x_{ei}=0.055$ \citep[obtained from RECFAST code\footnote{\url{http://www.astro.ubc.ca/people/scott/recfast.html}}][]{Seager1999, Seager2000}.

\subsection{Impact on heating due to the primordial magnetic field}
\label{subsec:effect-of-PMF}

The upper panel of Fig.~\ref{fig:PMF} shows the evolution of IGM temperature $T_g$ in presence of the primordial magnetic field with $B_0=0.05 \, {\rm nG}$ and $0.5 \, {\rm nG}$.  We fix $n_B = -2.9$ throughout our analysis. In order to understand the role of the primordial magnetic field alone we do not include the DM-baryon interaction in Fig.~\ref{fig:PMF}. Note that our results for $B_0=3 \, {\rm nG}$ and Hydrogen only scenario is very similar to that presented in \citet{Chluba15} for the similar scenario. We find that the primordial magnetic field makes a noticeable change in the IGM temperature during the cosmic dawn and dark ages ($z \lesssim 100$) for $B_0 \gtrsim 0.03 \, {\rm nG}$. This is because the ambipolar diffusion becomes very active at lower redshifts as  it is inversely proportional to the square of the baryon density, $\rho_b$. It also scales with the IGM temperature as $T^{-0.375}_{g}$ (see eq. \ref{eq:gamma_AD}).  The Effects due to the decaying turbulence, which scales as $\Gamma_{DT} \propto H(z) \, \rho_B(z)$, gets diluted at lower redshifts. We find that for $B_0 \sim 0.1 \, {\rm nG}$, the IGM temperature rises to the CMBR temperature and, consequently,  the global differential brightness temperature  $T_{21}$ becomes nearly zero. Further increase of the primordial magnetic field causes the IGM temperature goes above the CMBR temperature and $T_{21}$ becomes positive. This is completely ruled out as the EDGES measured the HI 21-cm signal in absorption i.e.,  $T_{21}$ is negative. This put an upper limit on the primordial magnetic field and we find  $B_0 \lesssim 0.1 \, {\rm nG}$, similar to the upper limit found by \citet{Minoda19}.

The middle panel of Fig.~\ref{fig:PMF} shows the history of residual free electron fraction, $x_e$. We see that $x_e$ increases if we increase the magnetic field $B_0$. This is because of  suppression in the Hydrogen recombination rate $\alpha_e$ due to increase in the IGM temperature $T_g$. The increase is more prominent during the cosmic dawn and dark ages. For example, $x_e$ increases by a factor of $\sim 1.5$ as compared to the standard prediction at redshift $z=17$ if $B_0=0.5 \, {\rm nG}$. Conversely, the residual free electron fraction $x_e$  directly influences the magnetic heating and its evolution through the ambipolar diffusion process (eq. \ref{eq:gamma_AD} and \ref{eq:mag_density}) which is dominant over the decaying turbulence during the cosmic dawn and dark ages. Therefore it is important to highlight the role of $x_e$ in constraining the primordial magnetic field using the global 21-cm signal. Moreover, $x_e$ also affects the standard Compton heating (see eq. \ref{eq:Tg}). 

The bottom panel of  Fig.~\ref{fig:PMF} shows the evolution of the primordial magnetic field. The normalised primordial magnetic field ($\frac{B(z)}{B_0(1+z)^2}$ ) has been plotted here to highlight any departure  from the simple $B_0(1+z)^2$ scaling. We find that the primordial normalised magnetic field maintains a constant value at higher redshifts $z \gtrsim 100$, and then decays at lower redshifts during the cosmic dawn and dark ages.  Because, a considerable fraction of the magnetic field energy is transferred to the IGM for its heating through the ambipolar diffusion process. The ambipolar diffusion becomes very active at lower redshifts for reasons explained in subsection \ref{subsec:PMF}. We also notice that the amount of decay of the magnetic field depends on $B_0$. For example, the normalised primordial magnetic field goes down to $\sim 0.4$ and $\sim 0.6$ for $B_0=0.05 \, {\rm nG}$ and $0.5 \, {\rm nG}$ at redshift $z \sim 17.2$. This implies that the fractional decay of the magnetic energy is more when the primordial magnetic filed is weaker. For higher primordial magnetic field with  $B_0 \gtrsim 1 \, {\rm nG}$, the fractional decay is not significant and it can be safely assumed to scale as $(1+z)^2$.

\subsection{Effect of dark matter-baryon interaction}
\label{subsec:effect of DM} 
\begin{figure}
	\includegraphics[width=\columnwidth]{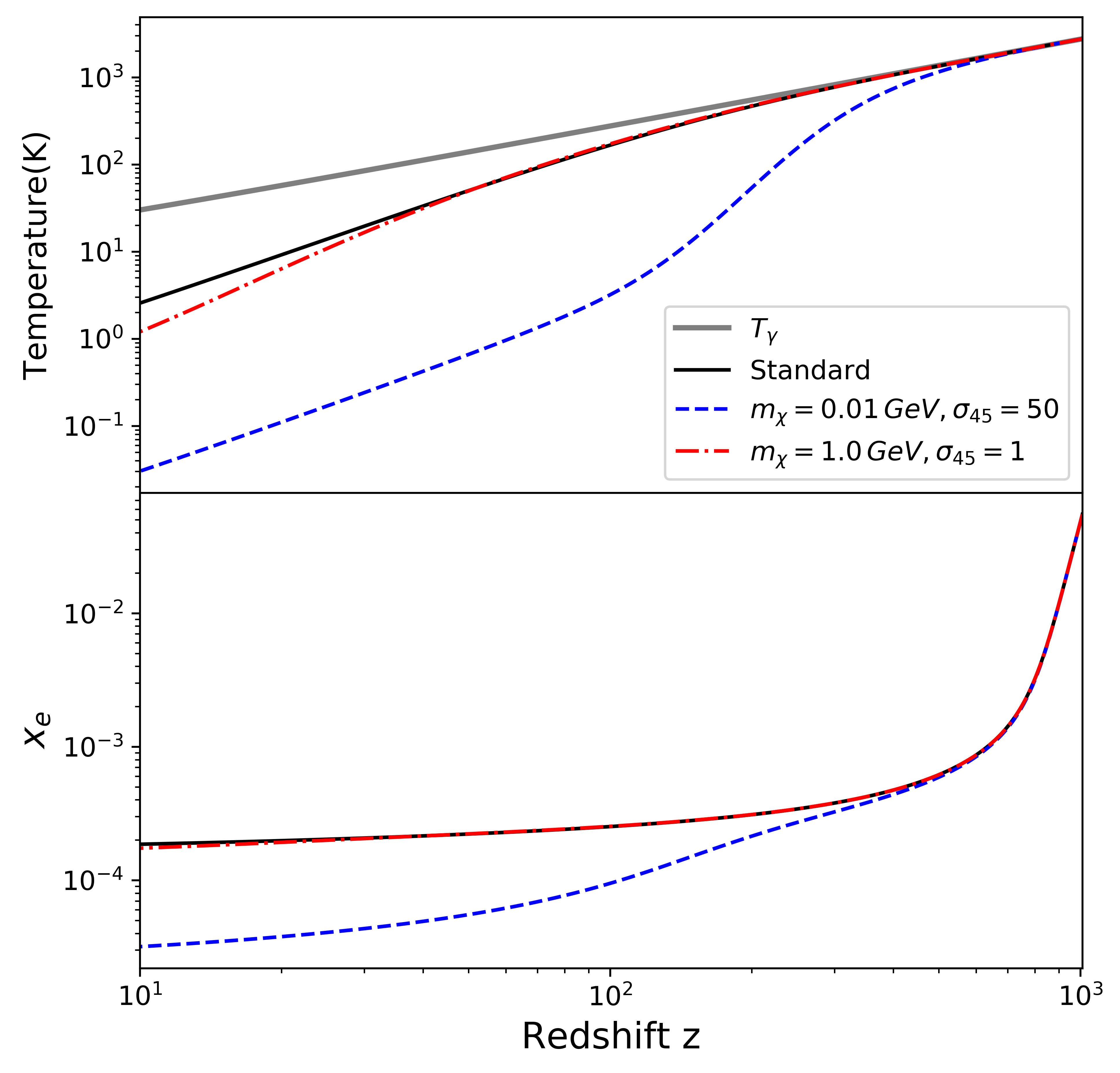}
    \caption{The upper and lower panels show the IGM kinetic temperature, $T_g$ and residual free electron fraction, $x_e$ as a function of redshift  when the DM-baryon interaction is considered.  The lower solid (black), dashed (blue) and dashed-dotted (red) lines correspond to  $(m_\chi/{\rm GeV}, \, \sigma_{45}) = (0,0)\, (0.01 , 50)$ and $(1 , 1)$ respectively. The effect due the primordial magnetic field is absent here. The upper solid (black) line represents the CMBR temperature $T_{\gamma}$.}
    \label{fig:DM}
\end{figure}

We consider the DM-baryon interaction model that was discussed in Sec.~\ref{subsec:dark-matter}. As mentioned there, the model has two free parameters i.e., the mass of the dark matter particle, $m_{\chi}$ and the interaction cross-section between the dark matter particles and baryons, $\sigma_{45}$. Below we briefly discuss the impact of the DM-baryon interaction on the IGM temperature, $T_g$ and residual free electron fraction, $x_e$. We refer readers to \citet{Datta20} for a more elaborate discussion.

The upper panel of Fig.~\ref{fig:DM} shows the evolution of IGM temperature for two sets of dark matter mass $m_\chi$ and interaction cross-section $\sigma_{45}$ i.e.,  $(1\, {\rm GeV},\, 1)$ and $(0.01\, {\rm GeV},\, 50)$. It also plots the IGM temperature as predicted in the standard model. As expected,  the interaction helps the IGM to cool faster and the IGM temperature becomes lower than the standard scenario during the cosmic dawn. Lower the dark matter mass, $m_\chi$ and/or larger the cross-section  $\sigma_{45}$, more is the rate of IGM cooling and, consequently, lower is the IGM temperature. We note that for higher cross section $\sigma_{45}$ the IGM temperature gets decoupled from the CMBR temperature early and coupled to the dark matter temperature $T_{\chi}$. This helps the IGM and the dark matter to reach the  thermal equilibrium. After that both the IGM and dark matter temperatures scale as $(1+z)^2$ which is seen at redshifts $z \lesssim 100$ for  $m_{\chi}=0.01\, {\rm GeV}$ and $\sigma_{45}=50$ (the blue-dashed curve in Fig. ~\ref{fig:DM}). Here we note that, there are mainly two effects arising due to the interaction between the cold DM and baryon. First, it helps to cool down the IGM faster (first term of rhs. of eq. \ref{eq:dQ_b}). Second, the friction due to the relative velocity between the DM and baryon can heat up both the DM and IGM (second term of rhs. of eq. \ref{eq:dQ_b})). We find that the friction heating dominates over the cooling for the DM particle mass $m_{\chi} \gtrsim 1\, {\rm GeV}$, and instead of cooling, the IGM gets heated due to the DM-baryon interaction for higher DM particle mass. However, in our case, we need faster cooling off the IGM. Therefore, the friction heating always remains subdominant in our case.  The bottom panel of Fig.~\ref{fig:DM} shows the  evolution of the residual free electron fraction, $x_e$.  As expected the residual free electron fraction is lower when the DM-baryon interaction comes into play. This is because the Hydrogen recombination rate $\alpha_e$ is increased when the IGM temperature is lower. The change in $x_e$ is not significant for $m_{\chi}=1$~GeV and $\sigma_{45}=1$ (red curve). However,   $x_e$  is reduced by factor of $\sim 5$  for $m_{\chi}=0.01\, {\rm GeV}$ and $\sigma_{45}=50$ (blue curve). The reduced $x_e$ enhances the rate of IGM heating through the ambipolar diffusion. At the same time lower IGM temperature reduces the coupling co-efficient $\gamma(T_g)$ (eq. \ref{eq:gamma_AD}), which again enhances the heating rate. Moreover, heating due to the Compton process, which is proportional to $x_e (T_{\gamma}-T_g)$ (second term on the rhs of eq. \ref{eq:Tg}),  gets affected when the IGM is colder compared to the standard scenario.

\begin{figure}
	\includegraphics[width=\columnwidth]{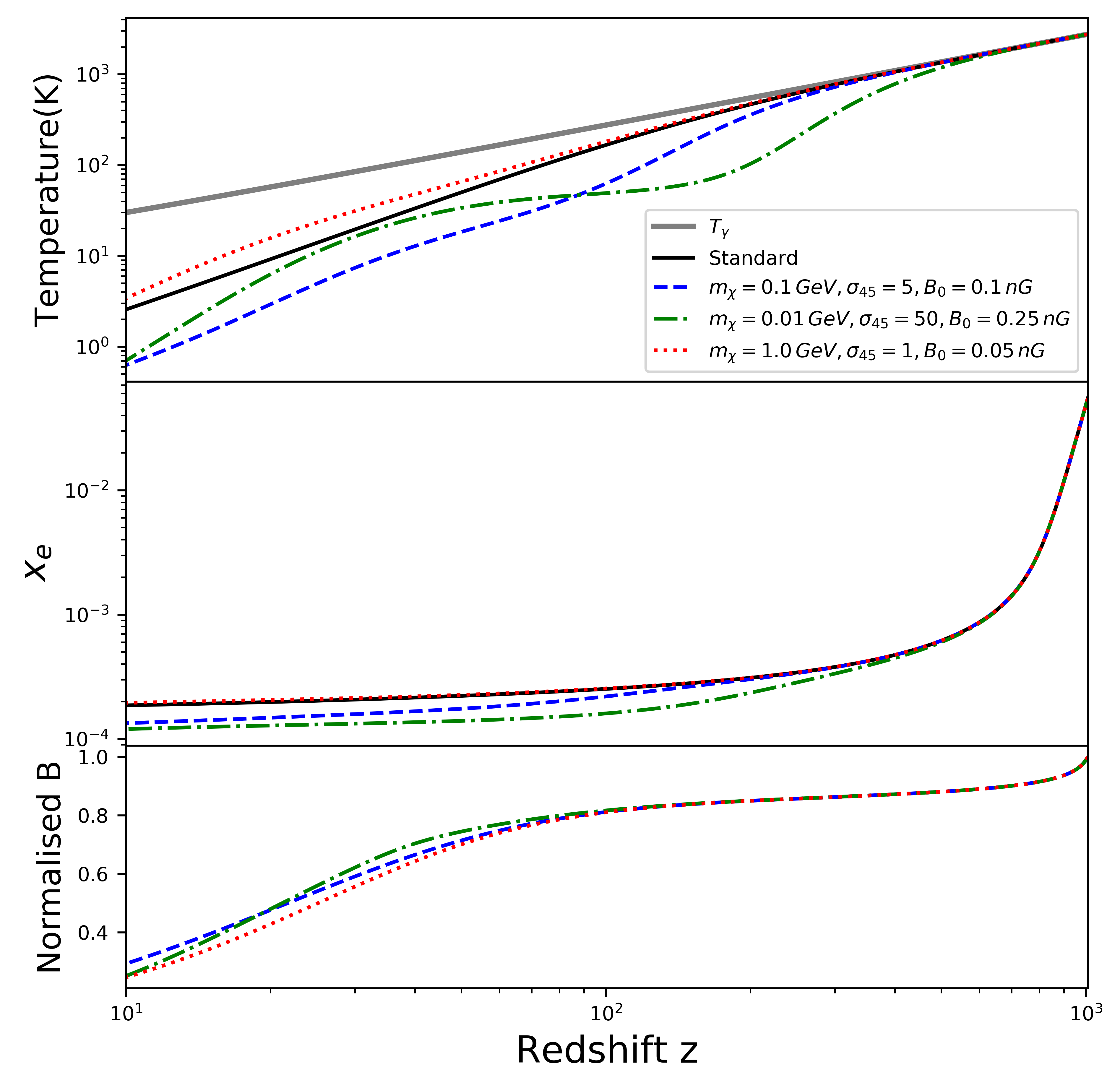}
    \caption{Same as Fig. \ref{fig:PMF}, however both the primordial magnetic field and DM-baryon interaction are considered here.}
    \label{fig:DM_PMF}
\end{figure}
\subsection{Combined impact of primordial magnetic field and dark matter-baryon interaction}
\label{subsec:effect of DM-PMF}
\begin{table}
	\centering
	\caption{The table shows the globally averaged differential brightness temperature $T_{21}$ at redshift $z=17.2$ for various set of the model parameters $m_{\chi}$, $\sigma_{45}$ and $B_0$. The allowed range of $T_{21}$ at redshift $z=17.2$ as measured by the EDGES is  $-0.3 \, {\rm K}$ to $-1.0 \, {\rm K}$.}
	\label{tab:table1}
	\begin{tabular}{lcccr} 
		\hline
		$m_\chi$ & $\sigma_{45}$ & $B_0$ & $T_{21}$ & Allowed\\
		$({\rm GeV})$ &  & $({\rm nG})$ & $({\rm K})$\\
		\hline
		$\times$ & $\times$ & $\times$ & -0.22 & $\times$\\
		$\times$ & $\times$ & 0.1 & 0.00 & $\times$\\
		0.1 & 5 & 0.1 & -0.87 & $\surd$\\
		1.0 & 1 & $\times$ & -0.62 & $\surd$\\
		1.0 & 1 & 0.05 & -0.15 & $\times$\\
		0.001 & 300 & 0.4 & -0.33 & $\surd$\\
		0.1 & 50 & 0.1 & -1.08 & $\times$\\
		\hline
	\end{tabular}
\end{table}

Here we discuss results on the combined impact of the primordial magnetic field and DM-baryon interaction on the IGM temperature evolution.  Fig. \ref{fig:DM_PMF} shows the evolution of the IGM temperature when both the  primordial magnetic field and DM-baryon interactions are considered. In Table~\ref{tab:table1} we have mentioned $T_{21}$ at $z=17.2$ as predicted by our models with different model parameters and shown which parameter set is allowed or not allowed by the EDGES measurements. We see that the differential brightness temperature $T_{21}$ at $z=17.2$,  for the parameter set $m_{\chi}=0.001\, {\rm GeV}$,  $\sigma_{45}=30$, is within the allowed range  when the primordial magnetic field with $B_0$ as high as $B_0=0.4 \, {\rm nG}$ is active, although $T_{21}$ is much lower when the magnetic field is kept off. Similarly, the parameter set $m_{\chi}=0.1\, {\rm GeV}$,  $\sigma_{45}=5$ is ruled out as it predicts much lower $T_{21}$ than what  is allowed by the EDGES data. However,  if we include the primordial magnetic field with, say, $B_0=0.1 \, {\rm nG}$, the above parameter set becomes allowed. Contrary to this, $T_{21}$ predicted by some combinations of $m_{\chi}$,  $\sigma_{45}$  could be well within the allowed range when there is no primordial magnetic field, but ruled out when the magnetic field is applied. For example $T_{21}=-0.62 \, {\rm K}$ for  $m_{\chi}=1 \, {\rm GeV}$,  $\sigma_{45}=1$ when $B_0=0$, but goes to $-0.15 \, {\rm K}$ which is above the allowed range for $B_0=0.05  \, {\rm nG}$.

We discussed in sub-section \ref{subsec:effect-of-PMF} that the primordial magnetic field with $B_0 \gtrsim 0.1  \, {\rm nG}$ is ruled out in the standard scenario, but it can be well within the allowed range when the interaction between DM and baryon with an appropriate parameter sets comes into play.  In general, we find that the exact upper limit on the primordial magnetic field depends on the mass of the DM particles $m_{\chi}$ and the DM-baryonic interaction cross section $\sigma_{45}$. We see that the primordial magnetic field with $B_0 \sim 0.4 \, {\rm nG}$ is allowed for an appropriate set of $m_{\chi}$ and $\sigma_{45}$. Note that this primordial magnetic field is ruled out in the standard scenario.

The upper panel of Fig. \ref{fig:DM_PMF} shows that the primordial magnetic field and DM-baryonic interaction together introduces a `plateau like feature' in the redshift evolution of the IGM temperature for a certain range of model parameters $m_{\chi}$, $\sigma_{45}$  and $B_0$. One such example can be seen for $m_{\chi}=0.01\, {\rm GeV}$,  $\sigma_{45}=50$  and $B_0=0.25 \, {\rm nG}$ where the plateau like feature is seen in redshift range $\sim 50-150$.  The cooling rate due to the DM-baryonic interaction and heating rate due to the primordial magnetic field compensates each other for a certain redshift range which gives the plateau like feature. At lower redshifts the heating due to the primordial magnetic field, which scales as $ B^4(z)$, becomes ineffective as the primordial magnetic field decays very fast. This is both due to the adiabatic expansion of universe and loss of the magnetic energy due to heating. We notice that this plateau like feature is not so prominent for lower primordial magnetic field. The `plateau like feature' is a unique signature of the DM-baryonic interaction in presence of the primordial magnetic field. However, it can only be probed by space based experiment as it appears at redshift range $\sim 50-150$.  

The middle and lower panels of Fig. \ref{fig:DM_PMF}  show the residual electron fraction, $x_e$ and primordial magnetic field, $B(z)$ as a function of redshift respectively. Like in previous cases, the residual electron fraction $x_e$ is suppressed  when both the DM-baryonic interactions and primordial magnetic field are active. The suppressed residual electron fraction enhances the heating rate occurring due to the ambipolar diffusion. The primordial magnetic  field looses its energy (other than the adiabatic loss because of universe's expansion) due to transfer of energy to IGM heating through the ambipolar diffusion process. This loss starts becoming important at lower redshifts $z \lesssim 100$.  As the primordial magnetic field decays very fast, the magnetic heating becomes ineffective at lower redshifts.  The EDGES absorption spectra show that the IGM temperature is rising  at  redshifts $z \lesssim 17$. There are several possible mechanisms by which the IGM can be heated up such as heating due to soft X-ray, Ly-$\alpha$, DM decay/annihilation \citep{pritchard2007, ghara14, ghara2020, sethi05, furlanetto2006, liu2018}. However, we find that the primordial magnetic field is not able to considerably heat up the IGM at the later phase of the cosmic dawn and, therefore, can not explain the heating part of the EDGES absorption profile. 

\subsection{Constraints on dark matter-baryon interaction in presence of the primordial magnetic field}
\label{subsec:DM-b-constraints}
\begin{figure*}
	\includegraphics[width=0.9\textwidth]{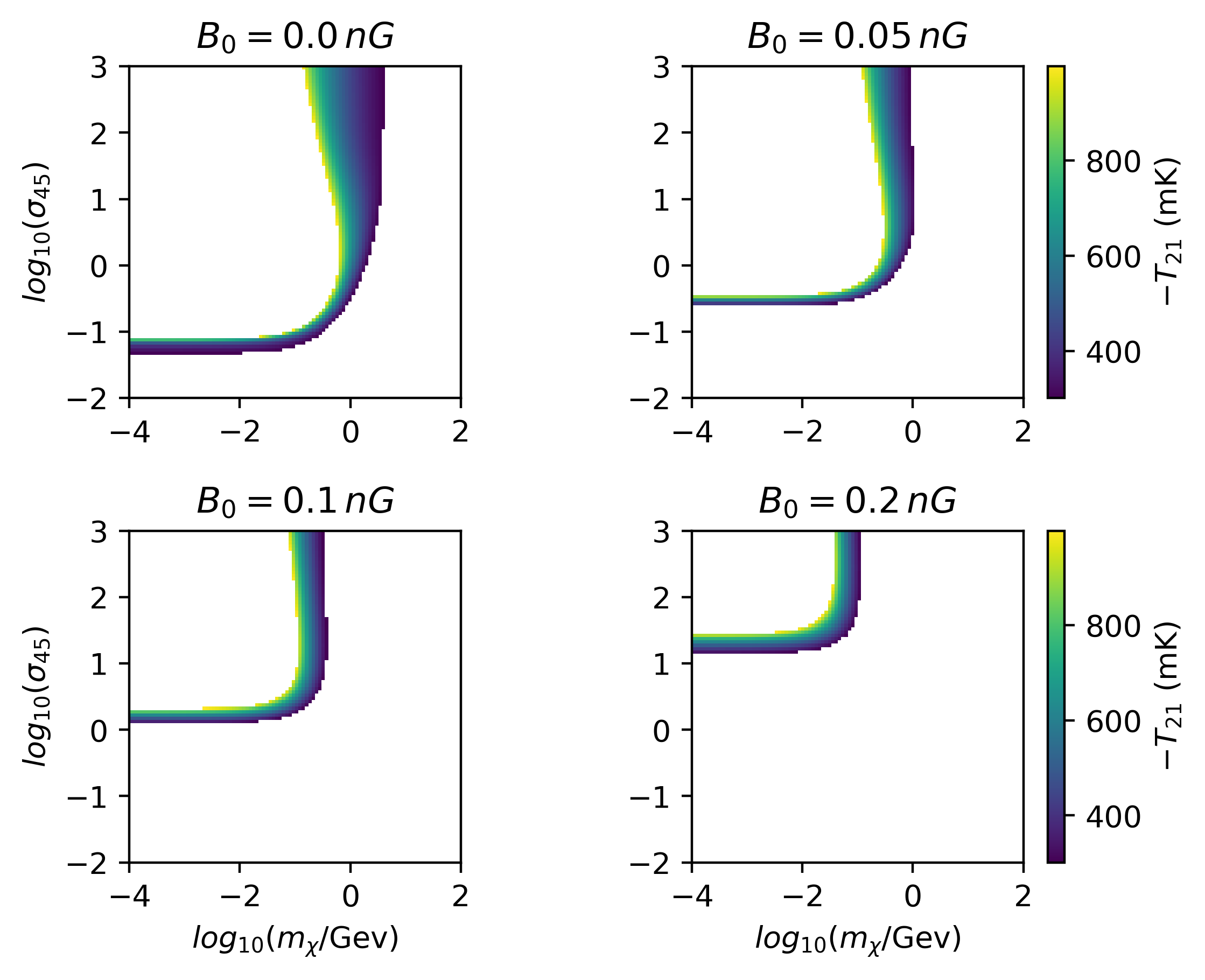}
    \caption{Bounds on dark matter mass and cross-section in presence of the primordial magnetic field.}
    \label{fig:contour}
\end{figure*}
Fig. \ref{fig:contour} demonstrates the  constraints on the DM-baryon interaction  in presence of the primordial magnetic field. The top left panel presents constraints on the model parameters $m_{\chi}$ and $\sigma_{45}$  when there is no magnetic field i.e., $B_0=0$. This is quite similar to constraints obtained by \citet{Barkana18Nature}.  Note that the constraints are obtained by restricting the differential brightness temperature $T_{21}$ within $-0.3 \, {\rm mK}$ to $-1.0 \, {\rm K}$ as suggested by the EDGES measurements.  The DM particle with mass higher than a few ${\rm GeV}$ is ruled out because the cooling due to the DM-baryonic interaction becomes inefficient and the drag heating due to the friction between the DM and baryon starts to dominate for higher DM particle mass. Therefore, the drag heating is found to have very negligible role in the case considered here. The top right, bottom left and the bottom right panels show constraints on the model parameters $m_{\chi}$ and $\sigma_{45}$  in presence of the primordial magnetic field with $B_0=0.05, \, 0.1 $ and $0.2 \, {\rm nG}$ respectively. We see that the allowed range of the DM-baryon cross section $\sigma_{45}$  gradually increases as $B_0$ is increased. For example, the lowest allowed $\sigma_{45}$  moves up, from $\sim 4 \times 10^{-47} \, {\rm m^2}$,  to $\sim 2.5 \times 10^{-46} \, {\rm m^2}$, $\sim 1.5 \times 10^{-45} \, {\rm m^2}$ and $\sim 1.5 \times 10^{-44} \, {\rm m^2}$ for $B_0=0.05, \, 0.1 $ and $0.2 \, {\rm nG}$ respectively. On the other hand, the maximum allowed mass of the DM particle $m_{\chi}$ gradually decreases for higher magnetic field. In Fig. \ref{fig:contour} we find that the highest allowed $m_{\chi}$ goes down, from $\sim 5  \, {\rm GeV}$, to $ \sim 1\, {\rm GeV}$,  $ \sim 0.3 \, {\rm GeV}$ and $\sim 0.1 \, {\rm GeV}$ for $B_0=0.05, \, 0.1 $ and $0.2 \, {\rm nG}$ respectively. The primordial magnetic field heats up the IGM and the heating is more for higher values of $B_0$. The DM-baryonic interaction needs to be more efficient  to compensates for this extra heating which can be achieved either by increasing the cross section $\sigma_{45}$ or/and lowering the mass of the Dark matter particle $m_{\chi}$.

The above discussion also tells that the exact upper limit on the primordial magnetic field parameter $B_0$ depends on the mass $m_{\chi}$ and the cross section $\sigma_{45}$. Higher primordial magnetic field is allowed if $\sigma_{45}$  is increased and/or  $m_{\chi}$ is decreased. We see that the primordial magnetic field with $B_0 \sim 0.4 \, {\rm nG}$ (Table \ref{tab:table1}) is allowed for an appropriate set of $m_{\chi}$ and $\sigma_{45}$. Note that $B_0 \gtrsim 0.1 \, {\rm nG}$ is ruled out in the standard scenario. However, we find that the primordial magnetic field with $B_0 \gtrsim 1 \, {\rm nG}$ may not be allowed as this requires very efficient cooling of the IGM which is unlikely even for very high cross section and lower DM particle mass. Although, we note that a recent study by \citet{bhatt2020}, which has used the EDGES measurements, finds an upper limit of $\sim 10^{-6} \, {\rm G}$ on the primordial magnetic field  for $m_{\chi} \lesssim 10^{-2} \, {\rm GeV}$ in presence of the DM-baryonic interaction.
 
\section{Summary and Discussion}
\label{sec:conclusion}
We study prospects of constraining the primordial magnetic field in light of the EDGES low band 21-cm absorption spectra during the cosmic dawn. Our analysis is carried out on the background of `colder IGM'  which is a promising avenue to explain the strong absorption signal found by the EDGES. We consider an interaction between baryons and cold DM particles which makes the IGM colder than in the standard scenario. The primordial magnetic field heats up the IGM through the ambipolar diffusion and decaying turbulence which, in turn, influences the 21-cm differential brightness temperature. We highlight the role of the residual electron fraction. We also study constraints on the DM-baryon interaction in presence of the primordial magnetic field, features in the redshift evolution of IGM temperature. In addition, we study redshift evolution of the primordial magnetic field during dark ages and cosmic dawn. In particular, we focus on the departure from the simple adiabatic scaling of the primordial magnetic field ( i.e. $B(z) \propto (1+z)^2$ ) due to the transfer of magnetic energy to the IGM.

Studying the role of the primordial magnetic field on the background of colder IGM is important for several reasons. First, it suppresses the abundance of the residual free electron fraction $x_e$  \citep{Datta20} which, in turn, enhances the rate of IGM heating through the ambipolar diffusion. Second, the coupling coefficient between the ionised  and neutral components $\gamma(T_g)$ decreases with the IGM temperature, which again results in the increased heating rate  (see eq. \ref{eq:gamma_AD}). Third, the heating rate due to the Compton process, which is proportional to  $(T_{\gamma} -T_g)$ and $x_e$, too gets affected when the background IGM temperature $T_g$ is lower (eq. \ref{eq:Tg}).  We find that collectively all these effects make the heating rate due the magnetic field faster in the colder background in compare to the heating rate in the standard scenario. Consequently, the primordial magnetic field decays, with redshift, at much faster rate compared to the simple $(1+z)^2$ scaling during the dark ages and cosmic dawn. The decay is particularly significant for $B_0 \lesssim 0.5 \, {\rm nG}$ when the fractional change in the magnetic field due to the heating loss could be $\sim 50 \%$ or higher. This is unique in the colder IGM scenario.  

Next we find that the upper limit on the primordial magnetic field using the EDGES measurements is determined by the underlying non-standard cooling process, i.e., the DM-baryon interaction here.  Higher primordial magnetic field may be allowed when the underlying DM-baryon interaction cross section is higher and/or the DM particle mass is lower, i.e.,  the exact upper limit on $B_0$ depends on the DM mass and the interaction cross section. For example, the primordial magnetic filed with $B_0 \sim 0.4 \, {\rm nG}$  which is ruled out in the standard model \citep{Minoda19}, may be allowed if the DM-baryon interaction with   $m_{\chi}=0.01 \, {\rm GeV}$ and $\sigma_{45}=100$ is included. However, we find that the primordial magnetic field with $B_0 \gtrsim 1 \, {\rm nG}$ may not be allowed as this requires very efficient cooling of the IGM which is unlikely to occur  even for very strong possible DM-baryon interaction.

Furthermore, we observe that the primordial magnetic field and DM-baryonic interaction together introduces `a plateau like feature' in the redshift evolution of the IGM temperature for a certain range of model parameters $m_{\chi}$, $\sigma_{45}$  and $B_0$. The cooling rate due to the DM-baryonic interaction and heating rate due to the primordial magnetic field compensates each other for a certain redshift range which produces the plateau like feature. However, this kind of plateau is not prominent for lower primordial magnetic field with $B_0 \lesssim 0.1 \, {\rm nG}$.

The EDGES absorption spectra suggest that the IGM temperature has possibly  gone up from $\sim 3 \, {\rm K}$ at redshift $z \approx 16$ to $\sim 40 \, {\rm K}$ at redshift $z \approx 14.5$. There are several possible candidates such soft X-ray photons from the first generation of X-ray binaries, mini-quasars, high energy photons from DM-decay/annihilations, primordial magnetic field etc. which could heat up the IGM during the cosmic dawn. However, our study shows that the heating due the primordial magnetic field becomes very weak during the above redshift range. Because the magnetic energy density decreases very fast prior to the cosmic dawn both due to the adiabatic expansion of universe and the loss due to IGM heating. Therefore, it is unlikely that the primordial magnetic field contributes to the heating of the IGM during the late phase of the cosmic dawn as indicated by the EDGES measurements. 

Finally, we see that the allowed DM-baryon cross section $\sigma_{45}$ gradually shifts towards higher values as $B_0$ is increases. On the other hand, the allowed mass of the DM particle $m_{\chi}$ gradually decreases for higher values of the primordial magnetic field. Because, the DM-baryon interaction needs to be more efficient to compensate for the excess heating caused due to higher magnetic field, which can be achieved either by increasing the cross section or lowering the mass of the Dark matter particle.

There could be various other models of the DM-baryon interactions, for which the exact upper limit on the primordial magnetic field, and all other results  discussed above might change to some extent. However, the general conclusions regarding the role of the primordial magnetic field on a colder IGM background are likely to remain valid for any mechanism providing faster cooling off the IGM.

\section*{Acknowledgements}
AB acknowledges financial support from UGC, Govt. of India. KKD and SS acknowledge financial support from BRNS through a project grant (sanction no: 57/14/10/2019-BRNS). KKD thanks Somnath Bharadwaj for useful discussion. SS thanks Presidency University for the support through FRPDF grant.



\bibliographystyle{mnras}
\bibliography{refs} 

\begin{thebibliography}{}
\makeatletter
\relax
\def\mn@urlcharsother{\let\do\@makeother \do\$\do\&\do\#\do\^\do\_\do\%\do\~}
\def\mn@doi{\begingroup\mn@urlcharsother \@ifnextchar [ {\mn@doi@}
  {\mn@doi@[]}}
\def\mn@doi@[#1]#2{\def\@tempa{#1}\ifx\@tempa\@empty \href
  {http://dx.doi.org/#2} {doi:#2}\else \href {http://dx.doi.org/#2} {#1}\fi
  \endgroup}
\def\mn@eprint#1#2{\mn@eprint@#1:#2::\@nil}
\def\mn@eprint@arXiv#1{\href {http://arxiv.org/abs/#1} {{\tt arXiv:#1}}}
\def\mn@eprint@dblp#1{\href {http://dblp.uni-trier.de/rec/bibtex/#1.xml}
  {dblp:#1}}
\def\mn@eprint@#1:#2:#3:#4\@nil{\def\@tempa {#1}\def\@tempb {#2}\def\@tempc
  {#3}\ifx \@tempc \@empty \let \@tempc \@tempb \let \@tempb \@tempa \fi \ifx
  \@tempb \@empty \def\@tempb {arXiv}\fi \@ifundefined
  {mn@eprint@\@tempb}{\@tempb:\@tempc}{\expandafter \expandafter \csname
  mn@eprint@\@tempb\endcsname \expandafter{\@tempc}}}

\bibitem[\protect\citeauthoryear{{Barkana}}{{Barkana}}{2018}]{Barkana18Nature}
{Barkana} R.,  2018, \mn@doi [\nat] {10.1038/nature25791}, \href
  {https://ui.adsabs.harvard.edu/abs/2018Natur.555...71B} {555, 71}

\bibitem[\protect\citeauthoryear{{Barkana}, {Outmezguine}, {Redigol}  \&
  {Volansky}}{{Barkana} et~al.}{2018}]{Barkana18PRD}
{Barkana} R.,  {Outmezguine} N.~J.,  {Redigol} D.,   {Volansky} T.,  2018,
  \mn@doi [\prd] {10.1103/PhysRevD.98.103005}, \href
  {https://ui.adsabs.harvard.edu/abs/2018PhRvD..98j3005B} {98, 103005}

\bibitem[\protect\citeauthoryear{{Bharadwaj} \& {Ali}}{{Bharadwaj} \&
  {Ali}}{2005}]{bharadwaj05}
{Bharadwaj} S.,  {Ali} S.~S.,  2005, \mnras, \href
  {http://adsabs.harvard.edu/abs/2005MNRAS.356.1519B} {356, 1519}

\bibitem[\protect\citeauthoryear{{Bhatt, Jitesh R.}, {Natwariya, Pravin Kumar},
  {Nayak, Alekha C.}  \& {Pandey, Arun Kumar}}{{Bhatt, Jitesh R.}
  et~al.}{2020}]{bhatt2020}
{Bhatt, Jitesh R.} {Natwariya, Pravin Kumar} {Nayak, Alekha C.}  {Pandey, Arun
  Kumar} 2020, \mn@doi [Eur. Phys. J. C] {10.1140/epjc/s10052-020-7886-x}, 80,
  334

\bibitem[\protect\citeauthoryear{{Bowman}, {Rogers}, {Monsalve}, {Mozdzen}  \&
  {Mahesh}}{{Bowman} et~al.}{2018}]{EDGES18}
{Bowman} J.~D.,  {Rogers} A. E.~E.,  {Monsalve} R.~A.,  {Mozdzen} T.~J.,
  {Mahesh} N.,  2018, \mn@doi [\nat] {10.1038/nature25792}, \href
  {https://ui.adsabs.harvard.edu/\#abs/2018Natur.555...67B} {555, 67}

\bibitem[\protect\citeauthoryear{{Chluba}, {Paoletti}, {Finelli}  \&
  {Rubi{\~n}o-Mart{\'\i}n}}{{Chluba} et~al.}{2015}]{Chluba15}
{Chluba} J.,  {Paoletti} D.,  {Finelli} F.,   {Rubi{\~n}o-Mart{\'\i}n} J.~A.,
  2015, \mn@doi [\mnras] {10.1093/mnras/stv1096}, \href
  {https://ui.adsabs.harvard.edu/abs/2015MNRAS.451.2244C} {451, 2244}

\bibitem[\protect\citeauthoryear{{Datta}, {Kundu}, {Paul}  \& {Bera}}{{Datta}
  et~al.}{2020}]{Datta20}
{Datta} K.~K.,  {Kundu} A.,  {Paul} A.,   {Bera} A.,  2020, arXiv e-prints,
  \href {https://ui.adsabs.harvard.edu/abs/2020arXiv200106497D} {p.
  arXiv:2001.06497}

\bibitem[\protect\citeauthoryear{{Furlanetto}, {Oh}  \&
  {Pierpaoli}}{{Furlanetto} et~al.}{2006a}]{furlanetto2006}
{Furlanetto} S.~R.,  {Oh} S.~P.,   {Pierpaoli} E.,  2006a, \mn@doi [\prd]
  {10.1103/PhysRevD.74.103502}, \href
  {https://ui.adsabs.harvard.edu/abs/2006PhRvD..74j3502F} {74, 103502}

\bibitem[\protect\citeauthoryear{{Furlanetto}, {Oh}  \& {Briggs}}{{Furlanetto}
  et~al.}{2006b}]{furlanetto06}
{Furlanetto} S.~R.,  {Oh} S.~P.,   {Briggs} F.~H.,  2006b, Physics Reports,
  \href {http://adsabs.harvard.edu/abs/2006PhR...433..181F} {433, 181}

\bibitem[\protect\citeauthoryear{{Ghara} \& {Mellema}}{{Ghara} \&
  {Mellema}}{2020}]{ghara2020}
{Ghara} R.,  {Mellema} G.,  2020, \mn@doi [\mnras] {10.1093/mnras/stz3513},
  \href {https://ui.adsabs.harvard.edu/abs/2020MNRAS.492..634G} {492, 634}

\bibitem[\protect\citeauthoryear{{Ghara}, {Choudhury}  \& {Datta}}{{Ghara}
  et~al.}{2015}]{ghara14}
{Ghara} R.,  {Choudhury} T.~R.,   {Datta} K.~K.,  2015, \mn@doi [\mnras]
  {10.1093/mnras/stu2512}, \href
  {http://adsabs.harvard.edu/abs/2015MNRAS.447.1806G} {447, 1806}

\bibitem[\protect\citeauthoryear{Jedamzik, Katalini\ifmmode~\acute{c}\else
  \'{c}\fi{}  \& Olinto}{Jedamzik et~al.}{1998}]{jedamzik98}
Jedamzik K.,  Katalini\ifmmode~\acute{c}\else \'{c}\fi{} V. c.~v.,   Olinto
  A.~V.,  1998, \mn@doi [Phys. Rev. D] {10.1103/PhysRevD.57.3264}, 57, 3264

\bibitem[\protect\citeauthoryear{{Kim}, {Olinto}  \& {Rosner}}{{Kim}
  et~al.}{1996}]{kim1996}
{Kim} E.-J.,  {Olinto} A.~V.,   {Rosner} R.,  1996, \mn@doi [\apj]
  {10.1086/177667}, \href
  {https://ui.adsabs.harvard.edu/abs/1996ApJ...468...28K} {468, 28}

\bibitem[\protect\citeauthoryear{{Kunze}}{{Kunze}}{2019}]{kunze2019}
{Kunze} K.~E.,  2019, \mn@doi [\jcap] {10.1088/1475-7516/2019/01/033}, \href
  {https://ui.adsabs.harvard.edu/abs/2019JCAP...01..033K} {2019, 033}

\bibitem[\protect\citeauthoryear{{Kunze} \& {Komatsu}}{{Kunze} \&
  {Komatsu}}{2014}]{KK14}
{Kunze} K.~E.,  {Komatsu} E.,  2014, \mn@doi [\jcap]
  {10.1088/1475-7516/2014/01/009}, \href
  {https://ui.adsabs.harvard.edu/abs/2014JCAP...01..009K} {2014, 009}

\bibitem[\protect\citeauthoryear{Liu \& Slatyer}{Liu \&
  Slatyer}{2018}]{liu2018}
Liu H.,  Slatyer T.~R.,  2018, \mn@doi [Phys. Rev. D]
  {10.1103/PhysRevD.98.023501}, 98, 023501

\bibitem[\protect\citeauthoryear{{Marinacci} \& {Vogelsberger}}{{Marinacci} \&
  {Vogelsberger}}{2016}]{marinacci2016}
{Marinacci} F.,  {Vogelsberger} M.,  2016, \mn@doi [\mnras]
  {10.1093/mnrasl/slv176}, \href
  {https://ui.adsabs.harvard.edu/abs/2016MNRAS.456L..69M} {456, L69}

\bibitem[\protect\citeauthoryear{{Minoda}, {Hasegawa}, {Tashiro}, {Ichiki}  \&
  {Sugiyama}}{{Minoda} et~al.}{2017}]{Minoda17}
{Minoda} T.,  {Hasegawa} K.,  {Tashiro} H.,  {Ichiki} K.,   {Sugiyama} N.,
  2017, \mn@doi [\prd] {10.1103/PhysRevD.96.123525}, \href
  {https://ui.adsabs.harvard.edu/abs/2017PhRvD..96l3525M} {96, 123525}

\bibitem[\protect\citeauthoryear{{Minoda}, {Tashiro}  \& {Takahashi}}{{Minoda}
  et~al.}{2019}]{Minoda19}
{Minoda} T.,  {Tashiro} H.,   {Takahashi} T.,  2019, \mn@doi [\mnras]
  {10.1093/mnras/stz1860}, \href
  {https://ui.adsabs.harvard.edu/abs/2019MNRAS.488.2001M} {488, 2001}

\bibitem[\protect\citeauthoryear{{Mu{\~n}oz} \& {Loeb}}{{Mu{\~n}oz} \&
  {Loeb}}{2018}]{Munoz18a}
{Mu{\~n}oz} J.~B.,  {Loeb} A.,  2018, \mn@doi [\nat]
  {10.1038/s41586-018-0151-x}, \href
  {https://ui.adsabs.harvard.edu/abs/2018Natur.557..684M} {557, 684}

\bibitem[\protect\citeauthoryear{{Mu{\~n}oz}, {Kovetz}  \&
  {Ali-Ha{\"\i}moud}}{{Mu{\~n}oz} et~al.}{2015}]{Munoz15}
{Mu{\~n}oz} J.~B.,  {Kovetz} E.~D.,   {Ali-Ha{\"\i}moud} Y.,  2015, \mn@doi
  [\prd] {10.1103/PhysRevD.92.083528}, \href
  {https://ui.adsabs.harvard.edu/abs/2015PhRvD..92h3528M} {92, 083528}

\bibitem[\protect\citeauthoryear{{Mu{\~n}oz}, {Dvorkin}  \& {Loeb}}{{Mu{\~n}oz}
  et~al.}{2018}]{Munoz18}
{Mu{\~n}oz} J.~B.,  {Dvorkin} C.,   {Loeb} A.,  2018, \mn@doi [\prl]
  {10.1103/PhysRevLett.121.121301}, \href
  {https://ui.adsabs.harvard.edu/abs/2018PhRvL.121l1301M} {121, 121301}

\bibitem[\protect\citeauthoryear{{Pandey}, {Choudhury}, {Sethi}  \&
  {Ferrara}}{{Pandey} et~al.}{2015}]{pandey2015}
{Pandey} K.~L.,  {Choudhury} T.~R.,  {Sethi} S.~K.,   {Ferrara} A.,  2015,
  \mn@doi [\mnras] {10.1093/mnras/stv1055}, \href
  {https://ui.adsabs.harvard.edu/abs/2015MNRAS.451.1692P} {451, 1692}

\bibitem[\protect\citeauthoryear{{Peebles}}{{Peebles}}{1968}]{Peebles1968}
{Peebles} P.~J.~E.,  1968, \mn@doi [\apj] {10.1086/149628}, \href
  {https://ui.adsabs.harvard.edu/abs/1968ApJ...153....1P} {153, 1}

\bibitem[\protect\citeauthoryear{{Planck Collaboration} et~al.,}{{Planck
  Collaboration} et~al.}{2016a}]{planck15}
{Planck Collaboration} et~al., 2016a, \mn@doi [\aap]
  {10.1051/0004-6361/201525830}, \href
  {https://ui.adsabs.harvard.edu/abs/2016A&A...594A..13P} {594, A13}

\bibitem[\protect\citeauthoryear{{Planck Collaboration} et~al.,}{{Planck
  Collaboration} et~al.}{2016b}]{planck15b}
{Planck Collaboration} et~al., 2016b, \mn@doi [A\&A]
  {10.1051/0004-6361/201525821}, 594, A19

\bibitem[\protect\citeauthoryear{{Pritchard} \& {Furlanetto}}{{Pritchard} \&
  {Furlanetto}}{2007}]{pritchard2007}
{Pritchard} J.~R.,  {Furlanetto} S.~R.,  2007, \mn@doi [\mnras]
  {10.1111/j.1365-2966.2007.11519.x}, \href
  {https://ui.adsabs.harvard.edu/abs/2007MNRAS.376.1680P} {376, 1680}

\bibitem[\protect\citeauthoryear{{Saga}, {Tashiro}  \& {Yokoyama}}{{Saga}
  et~al.}{2020}]{saga20}
{Saga} S.,  {Tashiro} H.,   {Yokoyama} S.,  2020, \mn@doi [\jcap]
  {10.1088/1475-7516/2020/05/039}, \href
  {https://ui.adsabs.harvard.edu/abs/2020JCAP...05..039S} {2020, 039}

\bibitem[\protect\citeauthoryear{Schleicher, Banerjee  \& Klessen}{Schleicher
  et~al.}{2008}]{Schleicher08}
Schleicher D. R.~G.,  Banerjee R.,   Klessen R.~S.,  2008, \mn@doi [Phys. Rev.
  D] {10.1103/PhysRevD.78.083005}, 78, 083005

\bibitem[\protect\citeauthoryear{{Schleicher}, {Banerjee}  \&
  {Klessen}}{{Schleicher} et~al.}{2009}]{Schleicher09}
{Schleicher} D. R.~G.,  {Banerjee} R.,   {Klessen} R.~S.,  2009, \mn@doi [\apj]
  {10.1088/0004-637X/692/1/236}, \href
  {https://ui.adsabs.harvard.edu/abs/2009ApJ...692..236S} {692, 236}

\bibitem[\protect\citeauthoryear{{Seager}, {Sasselov}  \& {Scott}}{{Seager}
  et~al.}{1999}]{Seager1999}
{Seager} S.,  {Sasselov} D.~D.,   {Scott} D.,  1999, \mn@doi [\apjl]
  {10.1086/312250}, \href
  {https://ui.adsabs.harvard.edu/abs/1999ApJ...523L...1S} {523, L1}

\bibitem[\protect\citeauthoryear{{Seager}, {Sasselov}  \& {Scott}}{{Seager}
  et~al.}{2000}]{Seager2000}
{Seager} S.,  {Sasselov} D.~D.,   {Scott} D.,  2000, \mn@doi [\apjs]
  {10.1086/313388}, \href
  {https://ui.adsabs.harvard.edu/abs/2000ApJS..128..407S} {128, 407}

\bibitem[\protect\citeauthoryear{{Sethi}}{{Sethi}}{2005}]{sethi05}
{Sethi} S.~K.,  2005, \mnras, \href
  {http://adsabs.harvard.edu/abs/2005MNRAS.363..818S} {363, 818}

\bibitem[\protect\citeauthoryear{{Sethi} \& {Subramanian}}{{Sethi} \&
  {Subramanian}}{2005}]{SS05}
{Sethi} S.~K.,  {Subramanian} K.,  2005, \mn@doi [\mnras]
  {10.1111/j.1365-2966.2004.08520.x}, \href
  {https://ui.adsabs.harvard.edu/abs/2005MNRAS.356..778S} {356, 778}

\bibitem[\protect\citeauthoryear{{Sethi}, {Nath}  \& {Subramanian}}{{Sethi}
  et~al.}{2008}]{sethi08}
{Sethi} S.~K.,  {Nath} B.~B.,   {Subramanian} K.,  2008, \mn@doi [\mnras]
  {10.1111/j.1365-2966.2008.13302.x}, \href
  {https://ui.adsabs.harvard.edu/abs/2008MNRAS.387.1589S} {387, 1589}

\bibitem[\protect\citeauthoryear{{Subramanian}}{{Subramanian}}{2016}]{kandu2016}
{Subramanian} K.,  2016, \mn@doi [Reports on Progress in Physics]
  {10.1088/0034-4885/79/7/076901}, \href
  {https://ui.adsabs.harvard.edu/abs/2016RPPh...79g6901S} {79, 076901}

\bibitem[\protect\citeauthoryear{Subramanian \& Barrow}{Subramanian \&
  Barrow}{1998}]{kandu98}
Subramanian K.,  Barrow J.~D.,  1998, \mn@doi [Phys. Rev. D]
  {10.1103/PhysRevD.58.083502}, 58, 083502

\bibitem[\protect\citeauthoryear{Takahashi, Mori, Ichiki, Inoue  \&
  Takami}{Takahashi et~al.}{2013}]{takahashi2013}
Takahashi K.,  Mori M.,  Ichiki K.,  Inoue S.,   Takami H.,  2013, \mn@doi [The
  Astrophysical Journal] {10.1088/2041-8205/771/2/l42}, 771, L42

\bibitem[\protect\citeauthoryear{{Tashiro} \& {Sugiyama}}{{Tashiro} \&
  {Sugiyama}}{2006}]{tashiro2006}
{Tashiro} H.,  {Sugiyama} N.,  2006, \mn@doi [\mnras]
  {10.1111/j.1365-2966.2006.10901.x}, \href
  {https://ui.adsabs.harvard.edu/abs/2006MNRAS.372.1060T} {372, 1060}

\bibitem[\protect\citeauthoryear{Tashiro, Kadota  \& Silk}{Tashiro
  et~al.}{2014}]{Tashiro14}
Tashiro H.,  Kadota K.,   Silk J.,  2014, \mn@doi [Phys. Rev. D]
  {10.1103/PhysRevD.90.083522}, 90, 083522

\bibitem[\protect\citeauthoryear{{Venumadhav}, {Oklop{\v{c}}i{\'c}},
  {Gluscevic}, {Mishra}  \& {Hirata}}{{Venumadhav}
  et~al.}{2017}]{Venumadhav2017}
{Venumadhav} T.,  {Oklop{\v{c}}i{\'c}} A.,  {Gluscevic} V.,  {Mishra} A.,
  {Hirata} C.~M.,  2017, \mn@doi [\prd] {10.1103/PhysRevD.95.083010}, \href
  {https://ui.adsabs.harvard.edu/abs/2017PhRvD..95h3010V} {95, 083010}

\bibitem[\protect\citeauthoryear{{Yamazaki}, {Ichiki}, {Umezu}  \&
  {Hanayama}}{{Yamazaki} et~al.}{2006}]{yamazaki2006}
{Yamazaki} D.~G.,  {Ichiki} K.,  {Umezu} K.-I.,   {Hanayama} H.,  2006, \mn@doi
  [\prd] {10.1103/PhysRevD.74.123518}, \href
  {https://ui.adsabs.harvard.edu/abs/2006PhRvD..74l3518Y} {74, 123518}

\makeatother
\end{thebibliography}





\bsp	
\label{lastpage}
\end{document}